# Comparing Machine Learning and Physics-Based Nanoparticle Geometry Determinations Using Far-Field Spectral Properties

Mengqi Sun,[†] Zixu Huang,[†] Muammer Y. Yaman,[†] Maxim Ziatdinov,[#,] Sergei V. Kalinin,[‡,#] David S. Ginger,[†, #, *]

[†]Department of Chemistry, University of Washington, Seattle, Washington, USA, 98195

[‡]Department of Materials Science and Engineering, University of Tennessee, Knoxville, TN 37996

[#] Physical Sciences Division Physical and Computational Sciences Directorate, Pacific Northwest National Laboratory, Richland, WA 99354, USA

*To whom correspondence should be addressed.

Email: dginger@uw.edu (David Ginger)

ABSTRACT

Anisotropic metal nanostructures exhibit polarization-dependent light scattering. This property has been widely exploited to determine geometries of subwavelength structures using far-field microscopy. Here, we explore the use of variational autoencoders (VAEs) to determine the geometries of gold nanorods (NRs) such as in-plane orientation and aspect ratio under linearly polarized dark-field illumination in an optical microscope. We input polarized dark-field scattering spectra and electron microscopy images into a dual-branch multimodal VAE with a single shared latent space trained on paired spectra–image data, using a learnable linear adapter. We achieve prediction of Au NRs using only polarized dark-field scattering spectra input. We determine geometrical parameters of orientational angle and aspect ratio quantitatively *via* both dual-VAE and physics-based analysis. We show that orientational angle prediction by dual-VAE performs well with only a small (~300 particle) training set, yielding a mean absolute error (MAE) of 14.4° and a concordance correlation coefficient (CCC) of 0.95. This performance is only marginally worse than the physics-based $\cos(2\theta)$ fitting approach between the scattering intensity and the polarizing angle, which achieves MAE of 8.78° and CCC of 0.99. Aspect ratio determination is also similar for the dual-VAE and physics-based fitting comparison (MAE of 0.21 *vs.* 0.23 and CCC of 0.53 *vs.* 0.68). By learning a shared latent manifold linking spectra and morphology, the model can generate NR images with accurate orientation and aspect ratio with spectra-only input in the small-data regime (~300 particles), suggesting a general recipe for inverse nano-optical problems requiring both structure and orientation information.

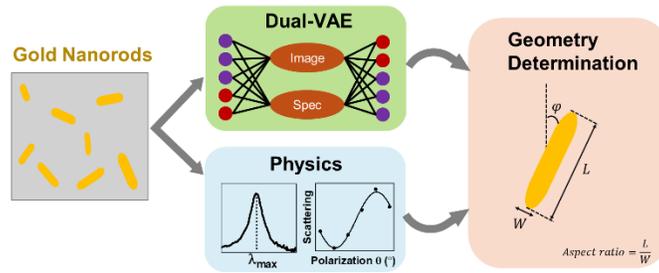

**TOC figure**

INTRODUCTION

The versatility of plasmonic nanoparticles and assemblies to interact strongly with light now finds applications in fields ranging from catalysis,[1-4] metamaterials,[5-7] and information display,[8,9] to chemical sensing,[10-14] and photothermal therapy.[15,16] The optical properties of these materials depend strongly on their shape, size, orientation and assembly into higher-order structures. Their strong structure-property relationship enables structural characterization using far-field information, which allows for rapid, non-invasive measurement of chemical and biological processes,[17-23] including nanoscale assembly and reaction dynamics.[24-28] Classical approaches to infer local geometries of plasmonic nanostructures often rely on direct measurements of known physical properties. In the case of probing orientational angles of anisotropic nanorods (NRs), polarization-resolved single-particle measurements, including dark-field scattering,[29-31] two-photon luminescence imaging,[32,33] photothermal imaging,[34] confocal microscopy,[35] and differential interference contrast imaging[36] are commonly reported in the literature.

More recently, researchers have applied machine learning (ML) models to capture the structural complexity of nanoscale assembly and orientation states.[37-40] For example, Hu et al. employed a differential interference contrast microscopy with a convolutional neural network to identify and predict both in-plane and out-of-plane orientation of individual metal nanoparticles, while achieving a full 0-360° range orientation detection using low-symmetry nanostars.[41] In addition, Long et al. used nonlinear ML that extracted low-dimensional subspaces from high-dimensional data to study the nonequilibrium self-assembly landscape of metallodielectric Janus particles under an applied AC electric field.[42] In another example, Newby and co-workers constructed convolutional neural network to automate particle tracking at the sub-micron scale.[43] Meanwhile, Yaman et al. developed dual variational autoencoders (dual-VAEs) to connect far-field scattering response with structural geometries of plasmonic clusters and effectively allowed predictions of optical response based on a known geometry, or reconstruction of assembled structures with a given scattering spectrum, albeit without orientational information.[27] Originally introduced in 2014,[44, 45] VAE is a probabilistic deep learning model that extracts meaningful characteristics of datasets in the encoders in a lower dimensional space (latent space) and generate predictions in the decoders. VAE has proven useful in addressing challenges of protein design,[46-48] molecular properties and structures discovery,[49-51] and image analysis.[52-55] VAE and other related deep learning ML models often excel in complex parameter spaces where physics-based analytical functions may not be available.[56, 57] Physics-based models on the other hand, when an empirical formula or analysis is available, can be faster and more readily-interpretable, but in other cases can require in-depth numerical solvers or *ab initio* computations.

Here, we evaluate the effectiveness of ML methods based on dual-VAEs to determine both the orientation and aspect ratio of anisotropic metal nanostructures using only far-field optical imaging. We study gold NRs as a model system and compare the performance of dual-VAEs trained with sparse datasets against analytical predictions using known physical relationships on real and heterogeneous samples of materials. We collect polarization-resolved dark-field scattering spectra of Au NRs and determine their in-plane orientation and aspect ratio using $\cos(2\theta)$ fitting[34, 58, 59] that captures the polarization direction-dependent scattering intensity and their scattering peak maxima, respectively. For the ML approach, we use dual-VAEs while enforcing a shared latent space, to enable orientational angle and aspect ratio determination *via* image reconstruction of Au NRs from the input angle-dependent spectra (**Fig. 1**). We compare both approaches and find that while cos(2θ) fitting delivers excellent performance in determining orientational angles of NRs with a mean absolute error (MAE) of 8.78° and Lin's concordance correlation coefficient (CCC) of 0.99, the dual-VAE approach also predicts angles well with a MAE of 14.44° and CCC of 0.95. Additionally, dual-VAE quantitatively determines the aspect ratio with a MAE of 0.21 and a CCC of 0.53, very comparable to the fitted aspect ratio *via* the scattering peak maxima as well (MAE of 0.23 and CCC of 0.68).

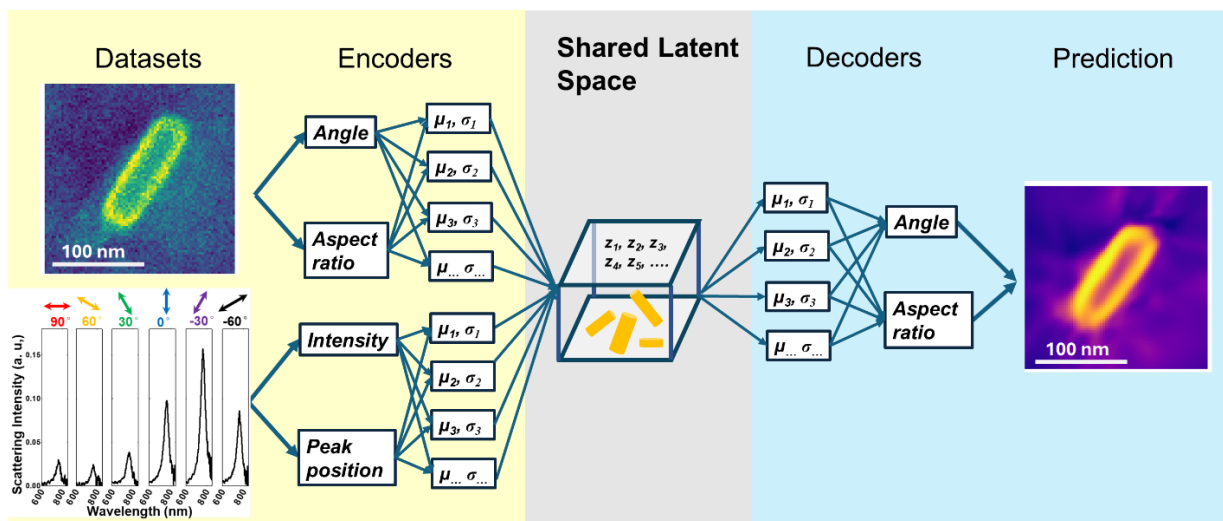

**Fig. 1.** Schematic illustration of dual-VAE approach with angle-resolved spectra. This encoder-decoder model encodes meaningful features of separate datasets (images and polarized spectra), enforces similarity at the shared latent space, and generates predictions in the decoders.

RESULTS AND DISCUSSION

We first synthesize Au NRs with aspect ratio of ~ 3 (118 nm × 38 nm) using a binary surfactant protocol.[60] Transmission electron microscopy (TEM) image (**Fig. S1**) shows the synthesized Au NRs to have good monodispersity. The extinction spectrum shows a transverse plasmonic peak at ~510 nm and longitudinal plasmonic peak at ~800 nm (**Fig. S2**). Using a dark-field scattering microscope equipped with linear polarizer at the incoming light pathway (**Fig. S3**), we collect linearly polarized scattering spectra of Au NRs at the single-particle level. While the optical imaging by the dark-field scattering is unable to resolve the geometrical details of the Au NRs, this information is encoded in their spectral properties, and correlating registered images with scanning electron microscopy (SEM) effectively allows visualization of the shape, size, and in-plane orientation of NRs (**Fig. 2A**). Upon designating 6 linear polarization directions, we collect spectra for the NRs and confirm that their longitudinal plasmonic scattering reaches a maximum when the light polarization is near-parallel to the long axis of NRs and minimum when near-orthogonal (**Fig. 2B**), as expected. This orientational dependence occurs because plasmonic resonance is maximized when surface free electron clouds oscillate in the same direction of incident light vector.[30, 34, 59, 61, 62] Spectra of other 3 NRs imaged in **Fig. 2A** are shown in **Fig. S4**. The brightness of dark-field scattering imaging of the diffraction-limited spots of each NRs also shows the orientational dependence (**Fig. S5**).

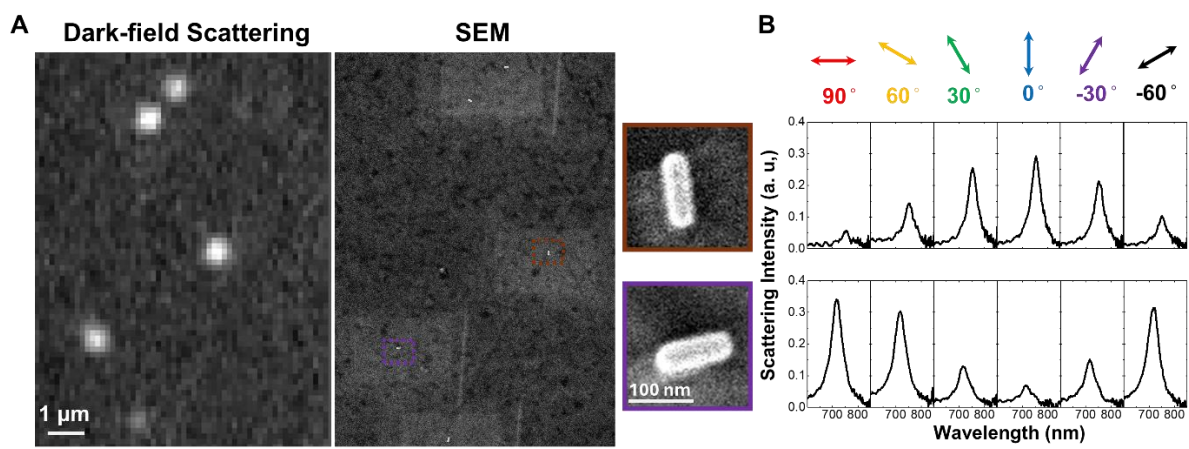

**Fig. 2.** (A) Correlated dark-field scattering and SEM image. Two NRs are highlighted. Their magnified views are shown. (B) Linearly polarized dark-field scattering spectra. Polarization directions are referenced and specified.

To illustrate the workflow of data training in the VAE, we outline each step in **Algorithm 1**, similar to our previous report.[27] During each training phase, encoder-1 processes the image data to generate latent vectors. These vectors are then fed into decoder-1 for a reconstruction of the original data. Encoder-2, on the other hand, takes dark-field scattering spectra data (All 6 linearly polarized scattering spectra combined into one array) to derive its latent representation. This representation undergoes a linear transformation before decoder-2 reconstructs it. The total loss function is given by equation (1):

$$\mathcal{L} = \mathcal{L}_{RE} + \beta \mathcal{L}_{KLD} + \gamma \mathcal{L}^1 \qquad (1)$$

Here, $\mathcal{L}_{RE}$ is a sum of reconstruction errors for decoded image and spectral data, $\mathcal{L}_{KLD}$ is a sum of Kullback-Leibler divergence (KLD) terms between standard normal distribution and encoded distributions of image and spectral data, and $\mathcal{L}^1$ is an L1 score between the latent representation of the encoder-1 and transformed latent representation of the encoder-2. The coefficients $\beta$ and $\gamma$ are constant scale factors that reflect the relative contribution of each term to the total loss function.

**Algorithm 1: Training of dual-VAE**

Inputs: Two experimental datasets, $X_1$ and $X_2$.

Pass $x_1$ through *encoder-1* to get $\mu_1$ and $\sigma_1$ parameters of variational distribution
Sample latent vector, $z_1 \sim \mathcal{N}(\mu_1, \sigma_1^2)$
Compute KL divergence, $D_1$, between encoded and prior distributions
Pass $z_1$ through *decoder-1* to obtain $x_1'$
Compute reconstruction loss, $RE_1$, between $x_1'$ and $x_1$

Pass $x_2$ through *encoder-2* to get $\mu_2$ and $\sigma_2$ parameters of variational distribution
Apply a learnable linear transformation, $A$, such that $\mu_2' = A\mu_2$.
Sample latent vector, $z_2' \sim \mathcal{N}(\mu_2', \sigma_2^2)$
Compute $L^1$ score between two latent vectors, $z_1$ and $z_2'$
Compute KL divergence, $D_2$, between encoded and prior distributions
Pass $z_2'$ through *decoder-2* to obtain $x_2'$
Compute reconstruction loss, $RE_2$, between $x_2'$ and $x_2$

Compute total loss, $\mathcal{L} = (RE_1 + RE_2) + \beta(D_1 + D_2) + \gamma L^1$
Backpropagate loss and adjust weights in both VAE models

At the prediction stage, we define a forward passing function that takes spectra data as the input through the spectra encoder (encoder-2) to obtain a latent vector which subsequently undergoes matrix projection in the shared latent space to get a transformed vector. The image decoder (decoder-1) therefore receives the transformed latent vector and reconstructs images of NRs.

We collect correlated dark-field scattering and SEM data of 320 individual NRs at various in-plane orientations and sizes for model training (**Fig. 3**) and use principal component analysis (PCA) method for batch processing the images to find orientational angle of each NRs (See Supporting Information for full workflow and **Fig. S6**). We process the polarization angle-dependent scattering spectra by simple concatenation of the 6 angle-dependent spectra into a single linear array (**Fig. 3B**), emphasizing the changes of scattering intensity as polarization direction varies.

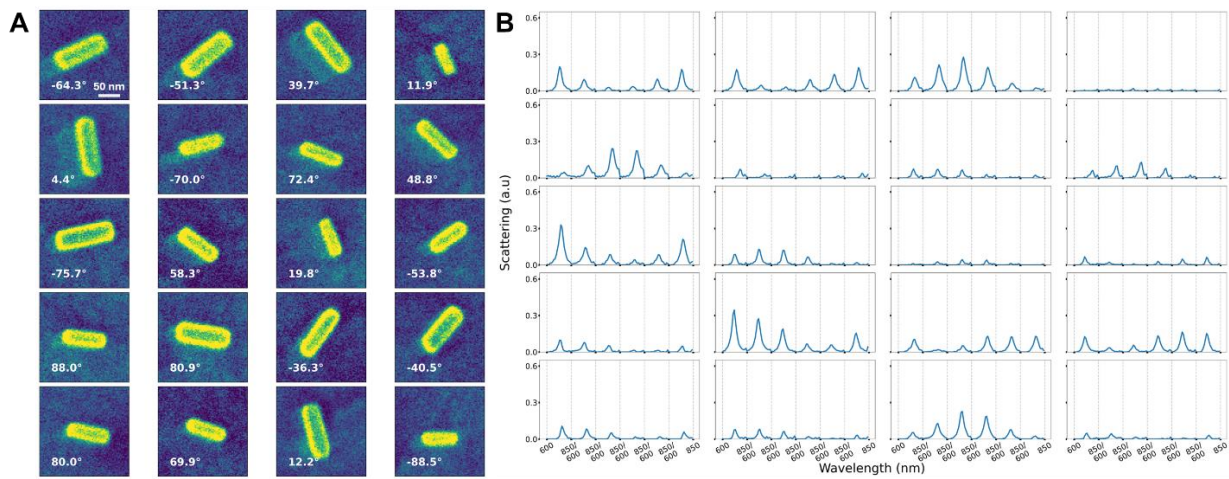

**Fig. 3.** Representative image and spectra dataset used for model training. (A) SEM images of individual NRs with in-plane orientational angles specified. (B) Dark-field scattering spectra of 6 linearly polarized spectra concatenated into one single linear array. The linear polarization direction of each spectrum from left to right is 90°, 60°, 30°, 0°, -30°, -60° respectively.

To evaluate how the latent space learns the relationships between the geometry and optical response of NRs. We first construct a dual-VAE model with only two latent vectors. The scatterplots of encoded latent representations, $Z_{2,image}$ *vs*. $Z_{1,image}$ and $Z_{2,spectra}$ *vs*. $Z_{1,spectra}$ show orientation-based clustering of NRs (**Fig. S7A and B**), consistent with the enforced similarity between the image and spectra VAEs during training. The decoded image latent space captures variations in the size and orientation of NRs (**Fig. S7C**), while the decoded spectral latent space captures variations in spectral properties, including relative scattering intensity and total intensity (**Fig. S7D**). However, we note that this 2D latent space does not appear to sort size and orientation separately, leading to relatively poor prediction performance on the length and width of NRs (**Fig. S8**).

We next designate three latent vectors and then revisit the scatterplots of the encoded latent representations (**Fig. S9**) and manifold representations of decoded latent space (**Fig. 4**). Similar to the 2D latent space, this 3D latent space captures the variations on the size and orientation of images data as well as the relative scattering intensity and total scattering intensity of spectra data. The two latent vectors, $Z_1$ and $Z_2$ seem to capture the orientational angle while $Z_3$ captures variation in the size of the NRs. At a given $Z_3$, the scatterplots of $Z_{2,image}$ *vs*. $Z_{1,image}$ and $Z_{2,spectra}$ *vs*. $Z_{1,spectra}$ demonstrate orientation-based clustering (**Fig. S10**). The dual-VAE does a better job learning aspect ratio correlations with 3 latent dimensions. **Fig. 4B** shows that, with increasing $Z_{3,image}$, the $Z_{2,image}$ *vs*. $Z_{1,image}$ latent slices of the decoded image show that length and width of NRs decrease while the decoded spectra show a corresponding decrease of the total scattering intensity as $Z_{3,spectra}$ increases. In other words, the correlated 3D latent space now appears to have learned a relationship between scattering intensity and nanoparticle size, which leads to improved performance in size prediction (**Fig. S11**).

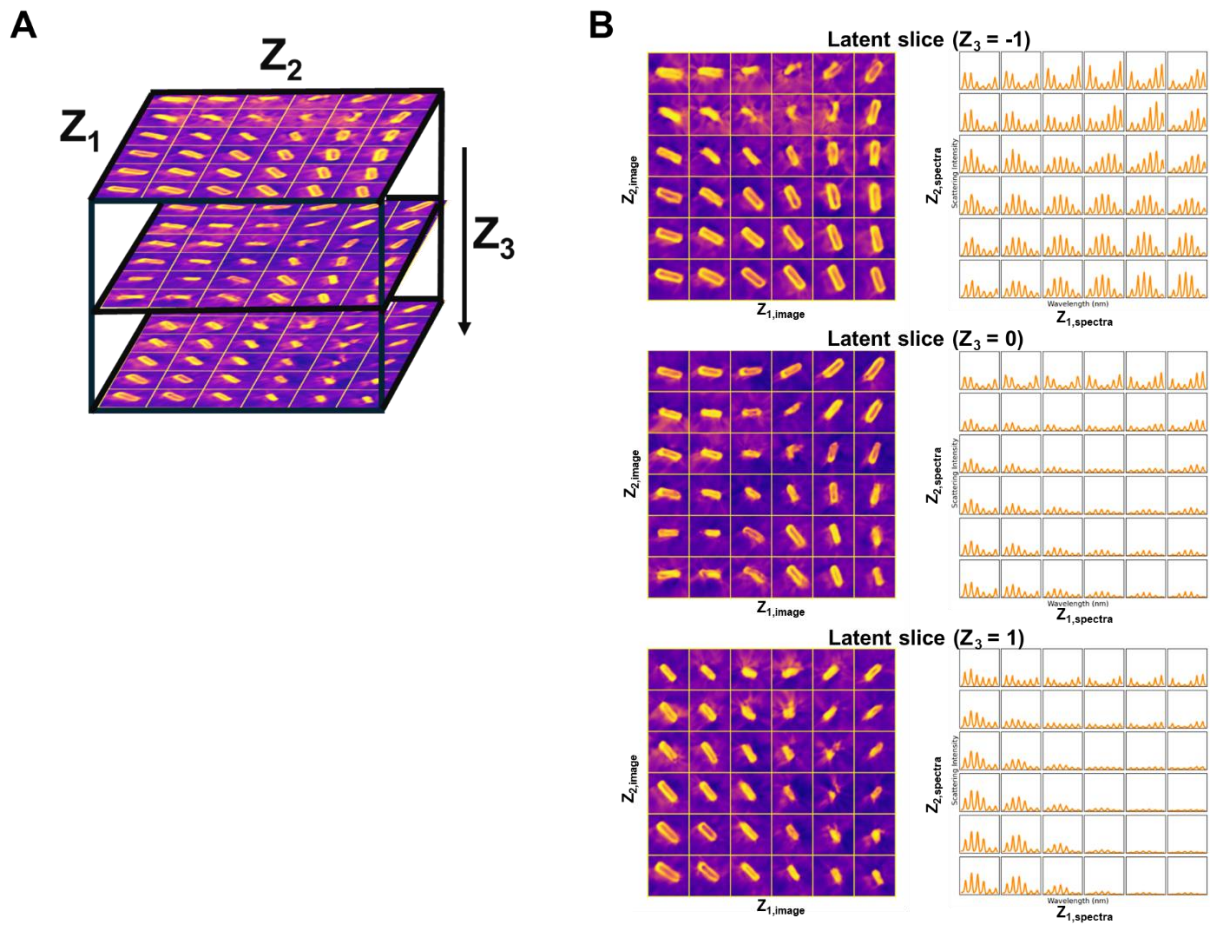

**Fig. 4.** (A) Schematics of 3-dimensional latent space manifold representation. (B) Latent slices of $Z_1$ *vs.* $Z_2$ manifold grids of decoded images and spectra at a given $Z_3$ that made up the 3-dimensional cubic latent space.

The enforced similarity of the two VAEs, manifested by the $\mathcal{L}^1$ score computation between latent vector from encoder-1 and transformed latent vector, is a crucial component of this dual encoder-decoder architecture. It allows cross-VAEs communication that enables the model to map meaningful features (relative intensity, total intensity, and peak position) of dark-field scattering spectra to the structural parameters (orientational angle, length, and width) of NRs. In **Fig. S12**, we see this enforced similarity between two VAEs resulting in a strong linear relationship between latent variables across VAEs ($Z_{1,spectra}$ *vs.* $Z_{1,image}$, $Z_{2,spectra}$ *vs.* $Z_{2,image}$, and $Z_{3,spectra}$ *vs.* $Z_{3,image}$).

Notably, this dual-VAE model shows comparable performance to a physics-based analytical method. We choose 40 NRs that are not included in the training dataset as the testing datasets. We then use their spectra data as input to be passed through the model for image prediction and the correlated image data as ground truth for the prediction performance evaluation. **Fig. 5A** shows representative ground truth and prediction images, demonstrating good agreement on shape, size, and in-plane orientation of NRs. We extract the ground truth geometrical parameters such as orientation angles and aspect ratios using image analysis of the SEM data. To perform the physics-based analysis, we use a conventional $\cos(2\theta)$ fitting approach to quantitatively determine the orientational angle of NRs *via* equation (2):

$$I(\theta) = N(1 + M\cos2(\theta - \varphi)) \qquad (2)$$

Here, $I$ is the scattering intensity, $\theta$ is the polarization angle, $N$ is the prefactor coefficient, $M$ is the modulation depth, and $\varphi$ is the orientational angle of the anisotropic image object (Au NRs). This $\cos(2\theta)$ fitting equation has been shown to precisely determine the orientational angle based on the polarization dependence.[34, 58, 59] **Fig. S13** shows a few representative $\cos(2\theta)$ fitting curves of the plasmonic scattering of NRs. We also use finite-difference time-domain (FDTD) computational method to simulate the polarization dependence of light scattering of NRs and perform $\cos(2\theta)$ fitting (see experimental section for details and **Fig. S14**) finding excellent agreement, as expected. Beyond extracting the in-plane orientational angle of the NRs from the $\cos(2\theta)$ fits using equation (2), the modulation depth $M$ quantifies the degree of polarization of a single NR's scattering. $M = 0$ corresponds to unpolarized scattering, while for an ideal linear dipole (*e.g.* a pure longitudinal plasmon of a rod), $M = 1$.[34] For our single-NR dataset, we find $< M > = 0.76 \pm 0.09$ on average (**Fig. S15A**). We explore the correlation between $M$ and aspect ratio, length, and width of NRs and find no clear relationship for our rods (**Fig. S15B-D**). While one might expect $M$ to contain shape information, over this range of the aspect ratios of NRs (~2.1-3.2), the modulation depth is not a good quantitative indicator for geometrical determination of aspect ratio due to the relatively large aspect ratio. To confirm this interpretation, we use FDTD to simulate the polarization-resolved light scattering of NRs with different aspect ratios from 1.0 (spherical particle) to 5.0 and perform $\cos(2\theta)$ fitting on each NRs (**Fig. S16A and B**). We see that the computed $M$ for ideal NRs increases sharply with aspect ratio but quickly plateaus (**Fig. S16C**) and is relatively insensitive to aspect ratio for aspect ratios of 2 or larger.

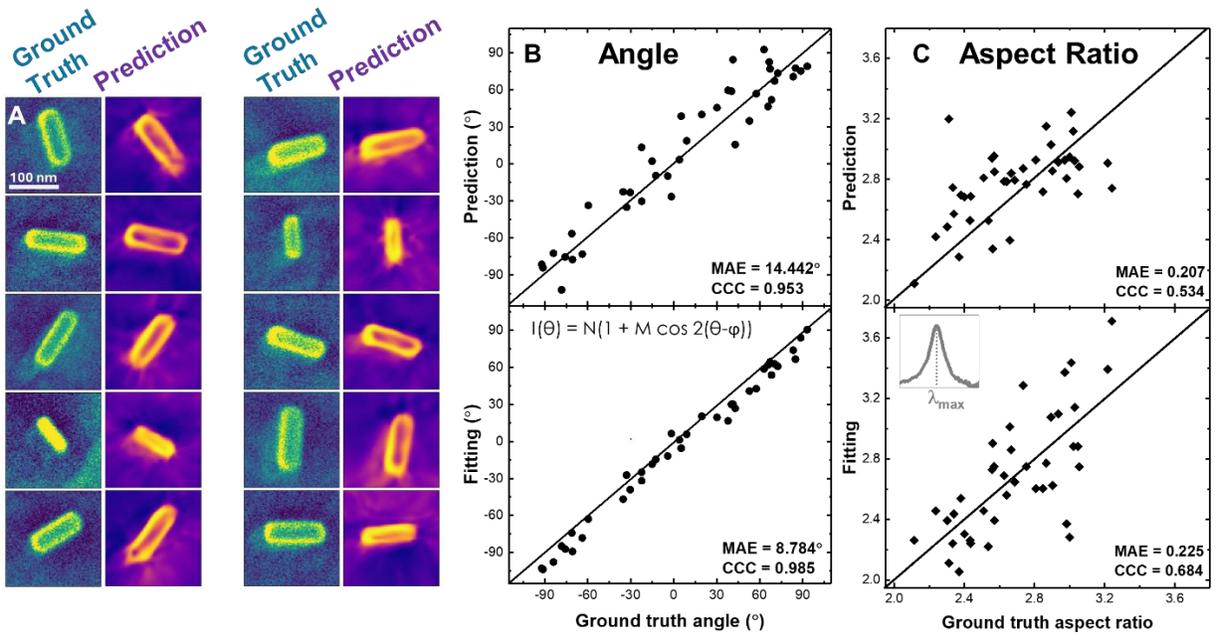

**Fig. 5.** Performance of orientation and aspect ratio determination using spectral properties comparing dual-VAE versus physics-based analytical approaches. (A) 10 representative ground truth and predicted images from the VAE. (B) Scatterplots of VAE-predicted angle and $\cos(2\theta)$-fit-determined angle (see **Fig. S13** for representative fits), (C) scatterplots of VAE-predicted aspect ratio and peak wavelength-fit-determined aspect ratio (**Fig. S18**).

To quantitively compare the performance of the dual-VAE and physics-based analyses to determine orientational angle and aspect ratio of NRs, we next plot both the predicted angles (dual-

VAE) and fit-determined angles (physics) versus the ground truth angles in **Fig. 5B**, along with the mean absolute error (MAE) and Lin's concordance correlation coefficient (CCC) as statistical metrics to evaluate the average absolute errors and alignment of prediction values with the identity line ($y = x$) respectively. We compute MAE as usual, *via* equation (3):

$$MAE = \frac{1}{n}\sum_{i=1}^{n}|\hat{y}_i - y_i| \tag{3}$$

Here, $\hat{y}_i$ is the predicted value, $y_i$ is the ground truth value, and n is the sample size. We compute CCC using equation (4):

$$CCC = \frac{2\rho\sigma_{\hat{y}_i}\sigma_{y_i}}{\sigma_{\hat{y}_i}^2 + \sigma_{y_i}^2 + (\mu_{\hat{y}_i} + \mu_{y_i})^2} \tag{4}$$

Here, $\rho$ is the Pearson's coefficient, $\sigma_{\hat{y}_i}$ and $\sigma_{y_i}$ are the standard deviation of predicted values and ground truth values, $\sigma_{\hat{y}_i}^2$ and $\sigma_{y_i}^2$ are the variances of predicted values and ground truth values, and $\mu_{\hat{y}_i}$ and $\mu_{y_i}$ are the means of predicted values and ground truth values. CCC has been used broadly in modeling and ML to assess how model outputs agree with the experimental observation.[63-66] Ranging from -1 to 1, a CCC of 1 indicates perfect agreement between prediction and observation, while at -1, CCC manifests a perfect disagreement or exact anticorrelation between prediction and observation.

Because the physics-based $\cos(2\theta)$ fitting approach effectively captures the strong $\cos(2\theta)$ dependence of plasmonic scattering response on the polarization direction, it gives a low MAE of 8.78° and a near-perfect CCC of 0.99. Notably however, the dual-VAE also produces good performance with a MAE of 14.44° and a marginally lower CCC, at 0.95 (this performance is almost identical relative to the 2D latent space dual VAE with a MAE of 14.58° and CCC of 0.95, illustrated by **Fig. S17**). On the other hand, we also plot the aspect ratio obtained from predicted images of NRs and fitting the peak position of scattering spectra against the ground truth (**Fig. 5C**). Since the longitudinal plasmonic peak of NRs is known to red shift with increasing aspect ratio,[29, 67] we apply a straightforward linear fit ($y = mx + b$) to capture the relationship between scattering peak position and aspect ratio (**Fig. S18**) to determine the fitted aspect ratio. We find that dual-VAE predicts aspect ratio with a similar performance compared to physics-based analysis. The average error magnitudes are slightly smaller for predicted aspect ratio by dual-VAE (MAE of 0.21 *vs.* 0.23) while fitted aspect ratio clusters around the identity line modestly better for physic-based fitting (CCC of 0.53 *vs.* 0.68) as the one notable outlier in aspect ratio prediction is penalized in CCC significantly. As noted above, 3D dual-VAE learns size of NRs significantly better than 2D dual-VAE as larger latent space leads to better size-based sorting of NRs. Therefore, the aspect ratio prediction performance by the 2D dual-VAE we discussed earlier is noticeably worse (MAE of 0.29 and CCC of only 0.32, illustrated by **Fig. S19**).

CONCLUSION

In conclusion, we demonstrate that dual-VAE as a generative ML model can effectively capture the relationships between geometrical information and polarization-resolved dark-field scattering of anisotropic Au NRs with only optical spectra as input and a relatively small training dataset (~300 NRs). We quantitatively determine geometrical parameters such as in-plane orientational angle and aspect ratio *via* image reconstruction with polarized scattering spectra as input. This dual-VAE predicts the orientational angle with good accuracy (14.44° in MAE) while regressing well to the ideal identity line ($y = x$) with a CCC of 0.95. Somewhat surprisingly, these numbers are comparable to the physics-based $\cos(2\theta)$ fitting approach that exploits the strong $\cos(2\theta)$ dependence of scattering intensity on polarization direction (MAE of 8.78° and CCC of 0.99).

Dual-VAE also predicts aspect ratio with good accuracy, showing an MAE of 0.21 that is slightly better than the physics-based approach (MAE of 0.23) where the peak position of the scattering spectra is used to find aspect ratio, although outliers skew the CCC for dual-VAE prediction slightly more harshly than for physics-based analysis (CCC of 0.53 *vs.* 0.68). The geometrical determination by dual-VAE reflects on the readiness and accuracy of prediction through image reconstruction of NRs, and the correlations in the latent spaces show that the VAE has effectively captured physical associations in the latent spaces. This work provides the groundwork to use similar VAE-based methods as orientation probes and for real-time or near real-time particle tracking, and for characterizing supramolecular assembly processes, taking advantage of the VAE's ability to perform well at inverse property-structure determination, including orientation, in the limit of sparse (few hundred particle) training sets.

EXPERIMENTS AND METHODS

**Chemicals and materials.** Cetyltrimethylammonium bromide (CTAB, 98.0%) and sodium oleate (NaOL, 97.0%) were purchased from TCI America. L-ascorbic acid (AA, 99.0%), chloroauric acid trihydrate ($3H_2O \cdot HAuCl_4$, 99.9 %), Sodium borohydride ($NaBH_4$, 99%), hydrochloric acid (HCl, 37 wt. %) and silver nitrate ($AgNO_3$, 99.9 %) were purchased from Sigma-Aldrich. $NaBH_4$ was stored in a nitrogen glovebox. Ultrapure water (18.2 MΩ) was used in all experiments. All chemicals were used as received without further purification.

**Characterizations.** The scanning electron microscopy (SEM) images were obtained using TFSApreo-S with Lovac Scanning Electron Microscope operating at 2 kV and 13 pA. Dark-field scattering spectra and microscopic images were collected on a Photon Etc. IMA system, using a Nikon Ni-U upright microscope and a 40× objective lens (Nikon, NA = 0.6). A standard tungsten halogen lamp was used as the illumination source. After passing through a linear polarizer (Thorlabs, Ø1") and transmitted dark-field condenser (T-CHA, Dry, NA = 0.95-0.80), incident light excited the NRs sample and scattered light was differentiated by the tunable volume Bragg grating filter with a step size of 2 nm and integration time of 2 seconds per step before collected by a CCD camera (Thorlabs, 1501M-US-TE), building a data cube by accumulating each wavelength slice from 600 to 850 nm. Scattering spectra were normalized by subtracting a dark spectrum near the spot of interest, followed by dividing by a white light reference spectrum using equation (5). White spectrum was collected daily.

$$I_{sample} = \frac{I_{as-collected} - I_{dark}}{I_{white}} \tag{5}$$

Transmission electron microscopy (TEM) images were obtained using The Tecnai G2 F20 Supertwin TEM microscope featuring field emission gun and operating at 200 kV.

**Synthesis of Au nanorods (NRs).** Following a literature protocol[60] with slight modifications, Au NRs with aspect ratio of 3.0 was synthesized. First, stock solution of 0.2 M CTAB, 0.01 M NaBH$_4$, 4 mM AgNO$_3$, 1 mM 3H$_2$O·HAuCl$_4$, 0.5 mM 3H$_2$O·HAuCl$_4$, and 0.064 M AA were prepared in water. Next, growth and seed solutions can be made separately. For the seed solution, 5 mL 0.2 M CTAB and 5 mL 0.5 mM 3H$_2$O·HAuCl$_4$ were mixed at room temperature before a subsequent injection of 0.6 mL of 0.01 M NaBH$_4$ under intensive magnetic stirring for 1 min. An instantaneous color change from yellow to brownish was observed upon injection. The seed solution was then left undisturbed for 30 mins before use.

For the growth solution, 0.7 g CTAB and 0.1234 g NaOL were dissolved in 25 mL water when heated to 60 ℃ for 30 mins. After the solution cooled down to 30 ℃, 2.4 mL of 4 mM AgNO$_3$ was added and left undisturbed for 15 mins. 25 mL of 1 mM 3H$_2$O·HAuCl$_4$ was next added under stirring at 700 rpm for 90 mins. A gradual color change from yellow to colorless was observed. Then 0.15 mL of concentrated HCl was added under stirring at 400 rpm for 15 mins, before 0.125 mL of 0.064 M AA and 20 µL of seed solution were added under stirring at 700 rpm for 1 min. Reaction solution was left undisturbed for 12 hours. A slight orange color was observed after 1 hour of reaction. Colloidal NRs solution was stored as synthesized. Before characterization, centrifugation-redispersion cycles were performed to remove excessive CTAB.

**Finite-difference time-domain (FDTD) simulations.** Finite-difference time-domain (FDTD) numerical simulations were performed using Ansys Lumerical 2021 R1.4 FDTD solver software. The simulation box was constructed with perfectly matched layer boundaries, mesh size of 5 nm, and a meshing override of 0.5 nm. Gold nanorod (120 nm in length and 35 nm in width) structure was modelled as one 85-nm long and 35-nm wide cylinder capped with two 35-nm diameter nanospheres using the optical properties of gold as reported by Johnson and Christy as modeled by König et al.[68, 69] A broadband total-field scattered-field (TFSF) source from 400 to 1000 nm at a step size of 2 nm was used as the illumination source. Scattered light was collected *via* frequency domain power monitors placed on each border on the exterior of the illumination source. The polarization angle of electric field vector varied from -90° to 90° with 15° increments. To test the modulation depth and its correlation with aspect ratio, we simulate NRs with various aspect ratio. We fix the width of the NRs at 40 nm while changing the length to be 48 nm, 60 nm, 80 nm, 120 nm, 160 nm, and 200 nm to tune the aspect ratio.

**Dual-VAEs**. The full VAE algorithms, methods, functions, and data processing techniques were executed on Juypter notebook, available for public access: https://github.com/GingerLabUW/NR_orientation_prediction_dualVAE. The image-spectra datasets and trained dual-VAE models are also shared in the repository. This notebook is largely based on our previously reported model.[27] Assisted by ChatGPT, notable revision and addition of algorithms include calculation of orientational angle of NRs using PCA, creating comparison figures between ground truth and predicted images, reformatting the spectra and image datasets.

ASSOCIATED CONTENT

Supporting Information is available: Additional TEM, extinction spectrum, dark-field scattering imaging, schematics of dark-field scattering microscopy, FDTD simulations, scatterplots of encoded latent representation, and more prediction results.

AUTHOR INFORMATION

**Corresponding Author**

David S. Ginger - Department of Chemistry, University of Washington, Seattle, Washington 98195, United States; Physical Sciences Division, Physical and Computational Sciences Directorate, Pacific Northwest National Laboratory, Richland, Washington 99352, United States; orcid.org/0000-0002-9759-5447  Email: dginger@uw.edu

**Authors**

Mengqi Sun - Department of Chemistry, University of Washington, Seattle, Washington 98195, United States; orcid.org/0000-0002-3500-4189

Zixu Huang - Department of Chemistry, University of Washington, Seattle, Washington 98195, United States

Muammer Y. Yaman - Department of Chemistry, University of Washington, Seattle, Washington 98195, United States; orcid.org/0000-0002-9146-8105

Maxim Ziatdinov - Physical Sciences Division, Pacific Northwest National Laboratory, Richland, Washington 99354, United States

Sergei V. Kalinin - Department of Materials Science and Engineering, University of Tennessee, Knoxville, Tennessee 37909, United States; Physical Sciences Division, Pacific Northwest National Laboratory, Richland, Washington 99354, United States; orcid.org/0000-0001-5354-6152

**Author contribution**

M.S., M.Y.Y. and D.S.G drafted the paper and conceived the idea. M.S. performed synthesis of Au NRs, collected SEM images and dark-field scattering spectra, revised and executed dual-VAE algorithms. Z.H wrote the algorithms for preprocessing images and spectra data. M.Z. and S.V.K. originally wrote the dual-VAE algorithms. All authors contributed to the manuscript writing.

**Notes**

Authors declare no competing financial interests

ACKNOWLEDGEMENT

This material is based primarily upon work supported by the US Department of Energy, Office of Science, Office of Basic Energy Sciences, as part of the Energy Frontier Research Centers program: CSSAS – The Center for the Science of Synthesis Across Scales under Award Number DE-SC0019288. SEM imaging was conducted at the University of Washington Molecular Analysis Facility, a National Nanotechnology Coordinated Infrastructure (NNCI) site which was supported in part by the National Science Foundation, the University of Washington, the Molecular Engineering and Sciences Institute, and the Clean Energy Institute. Maxim Ziatdinov acknowledges support from Oak Ridge National Laboratory's Center for Nanophase Materials Sciences (user proposal project number CNMS2021-B-00847), a U.S. Department of Energy, Office of Science User Facility, for the development of initial prototypes of the VAE codebase that were later adapted under this project. We also acknowledge use of the codebase where the unmodified VAE tools are made publicly available, a repository which is supported by Laboratory Directed Research and Development Program at Pacific Northwest National Laboratory, a multiprogram national laboratory operated by Battelle for the U.S. Department of Energy. We acknowledge the use of ChatGPT-V5 in proofreading the manuscript for grammar and readability and assisting with code generation as described in the methods, the authors take full responsibility for all text and results.

Supporting Information for

# Comparing Machine Learning and Physics-Based Nanoparticle Geometry Determinations Using Far-Field Spectral Properties

*Mengqi Sun, † Zixu Huang, † Muammer Y. Yaman, † Maxim Ziatdinov, #, Sergei V. Kalinin, ‡, # David S. Ginger,†, #, ***

†*Department of Chemistry, University of Washington, Seattle, Washington, USA, 98195*

‡*Department of Material Science and Engineering, University of Tennessee, Knoxville, TN 37996*

# *Physical Sciences Division Physical and Computational Sciences Directorate, Pacific Northwest National Laboratory, Richland, WA 99354, USA*

* *To whom correspondence should be addressed.*

 *Email: dginger@uw.edu (David Ginger)*

1. **Principal Component Analysis (PCA) to calculate orientational angles**

Principal Componenet Analysis (PCA) is a dimensionality reduction statistical model that is frequently used for image analysis and recognition.[1-4] As illustrated in the algorithms in **Fig. S6**, this method first converts the image to a simpler grayscale format before thresholding and masking, which separates the image objects from background pixels. Then, coordinate information of the image object is extracted and used by the PCA analysis to identify the principal axis (long axis of NRs) that has the maximum variances (pixels). Finally, an inverse tangent operation between the first principal component vector along the y-axis direction (eigenvector[1]) and the x-axis direction (eigenvector[0]) calculates the orientational angle relative to x-axis. We further shift all the angles of [90, 180] to [-90,0] to keep angle referencing consistent.

2. **Calculation of mean absolute error (MAE) and concordance correlation coefficient (CCC)**

Mean absolute error (MAE) measures the average of absolute prediction errors of each datapoint. It indicates the deviation magnitude and carries the same unit with the data. We calculate MAE via equation (**S1**):

$$MAE = \frac{1}{n}\sum_{i=1}^{n}|\hat{y}_i - y_i| \qquad (S1)$$

Here, $\hat{y}_i$ is the predicted value, $y_i$ is the ground truth value, and n is the sample size.

Concordance correlation coefficient (CCC) captures the alignment of datapoint with respect to the identity line ($y = x$) and is useful to identify scale error and bias. CCC is a dimensionless unit as it ranges from -1 to 1. $CCC = 1$ indicates perfect agreement while $CCC = -1$ means perfect disagreement. We compute CCC via equation (**S2**):

$$CCC = \frac{2\rho\sigma_{\hat{y}_i}\sigma_{y_i}}{\sigma_{\hat{y}_i}^2 + \sigma_{y_i}^2 + (\mu_{\hat{y}_i} + \mu_{y_i})^2} \qquad (S2)$$

Here, $\rho$ is the Pearson's coefficient, $\sigma_{\hat{y}_i}$ and $\sigma_{y_i}$ are the standard deviation of predicted values and ground truth values, $\sigma_{\hat{y}_i}^2$ and $\sigma_{y_i}^2$ are the variances of predicted values and ground truth values, and $\mu_{\hat{y}_i}$ and $\mu_{y_i}$ are the means of predicted values and ground truth values.

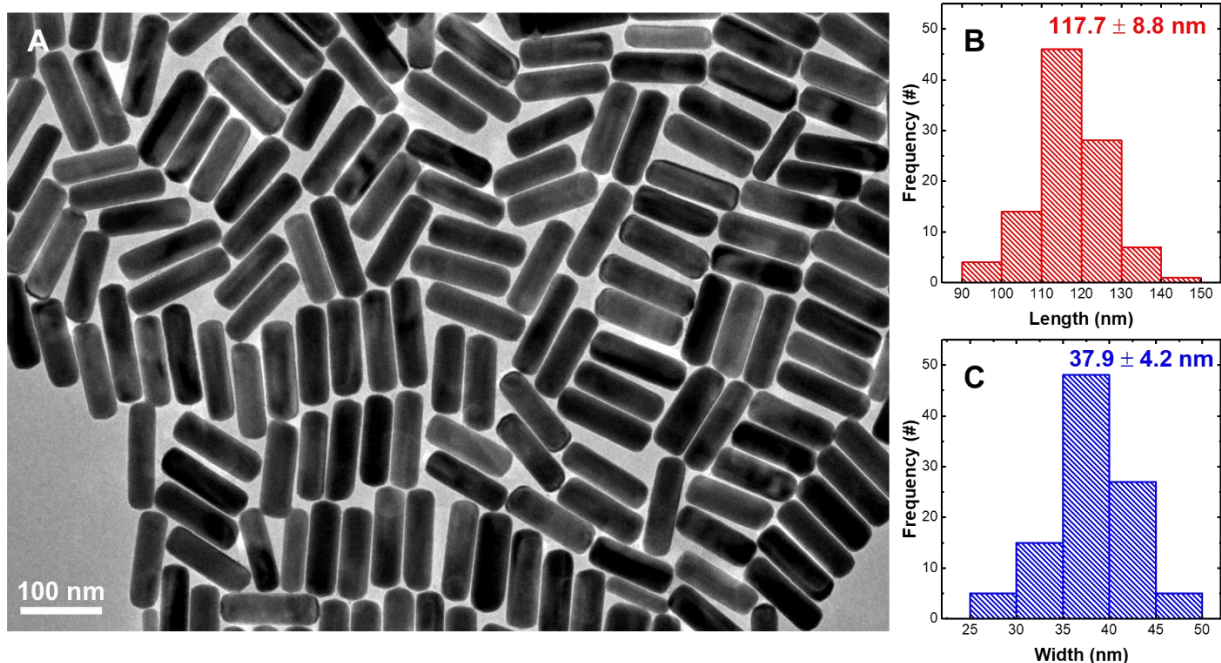

**Fig. S1.** (A) TEM image of synthesized NRs. Histograms of (B) length and (C) width of NRs. Based on the TEM image, length and width of NRs were determined to be 117.7 ± 8.8 nm and 37.9 ± 4.2 nm respectively.

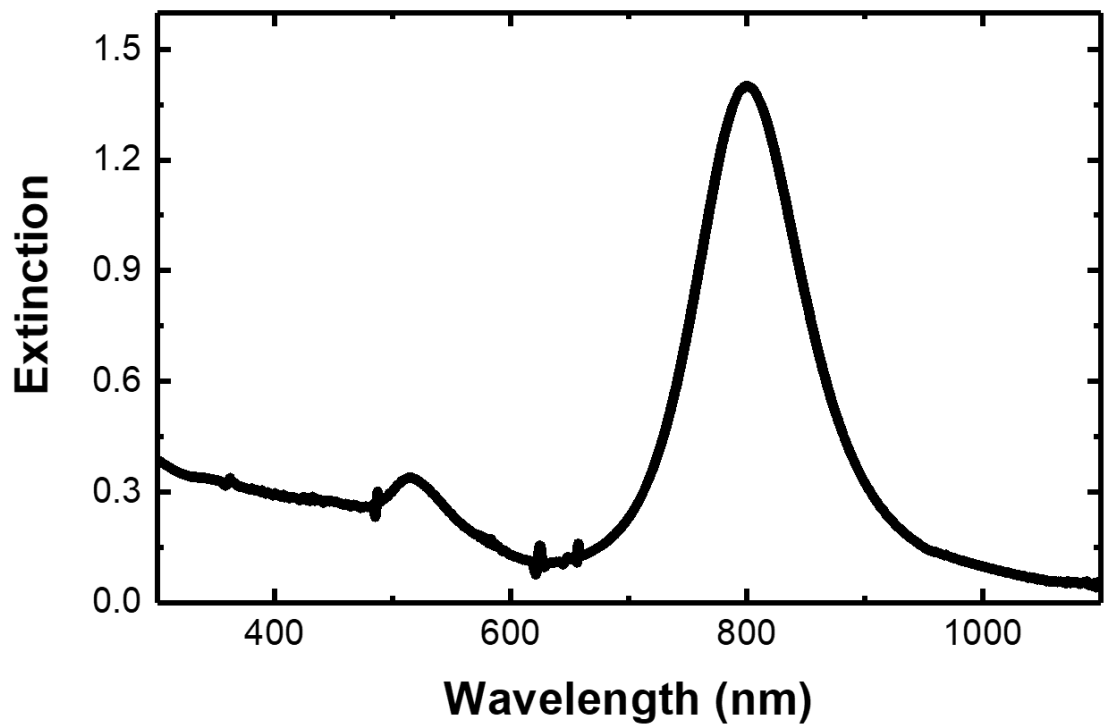

**Fig. S2.** Extinction spectrum of synthesized NRs.

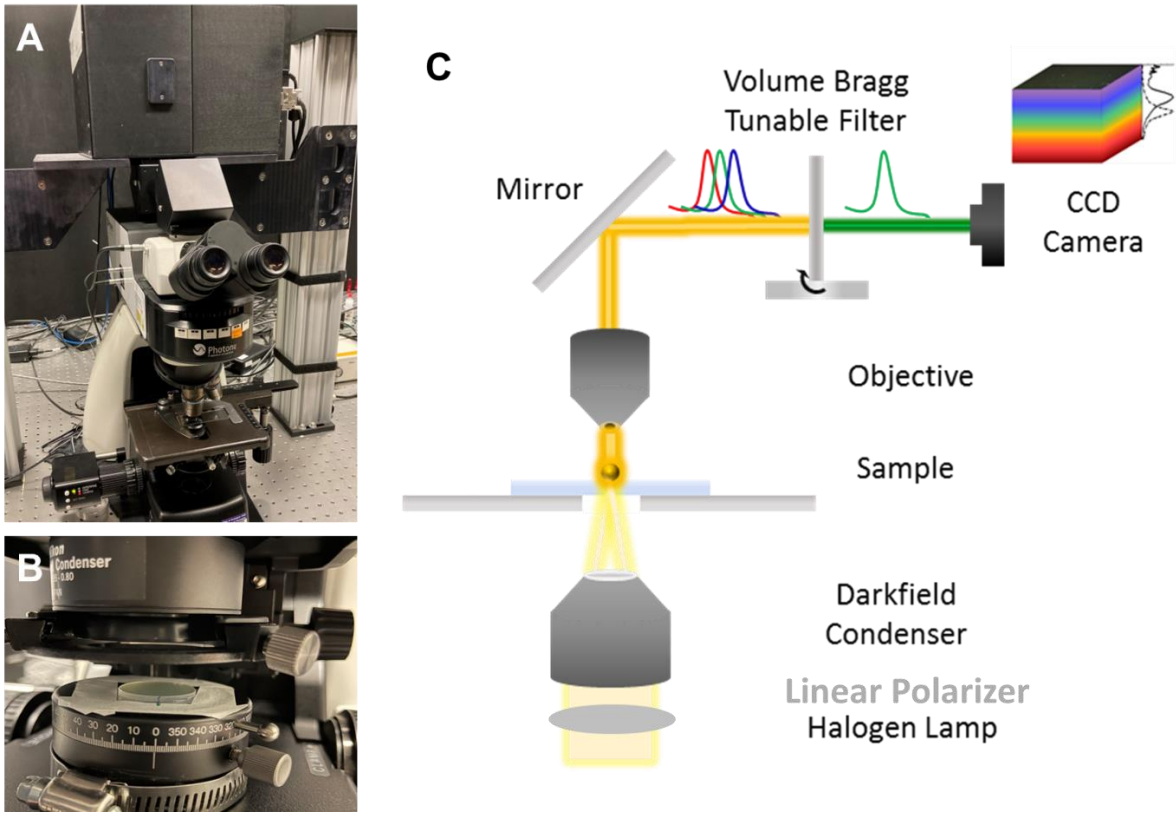

**Fig. S3.** Photographs of (A) the dark-field scattering microscope and (B) linear polarizers mounted below the darkfield condenser. (C) Schematic illustration of the dark-field scattering microscopy.

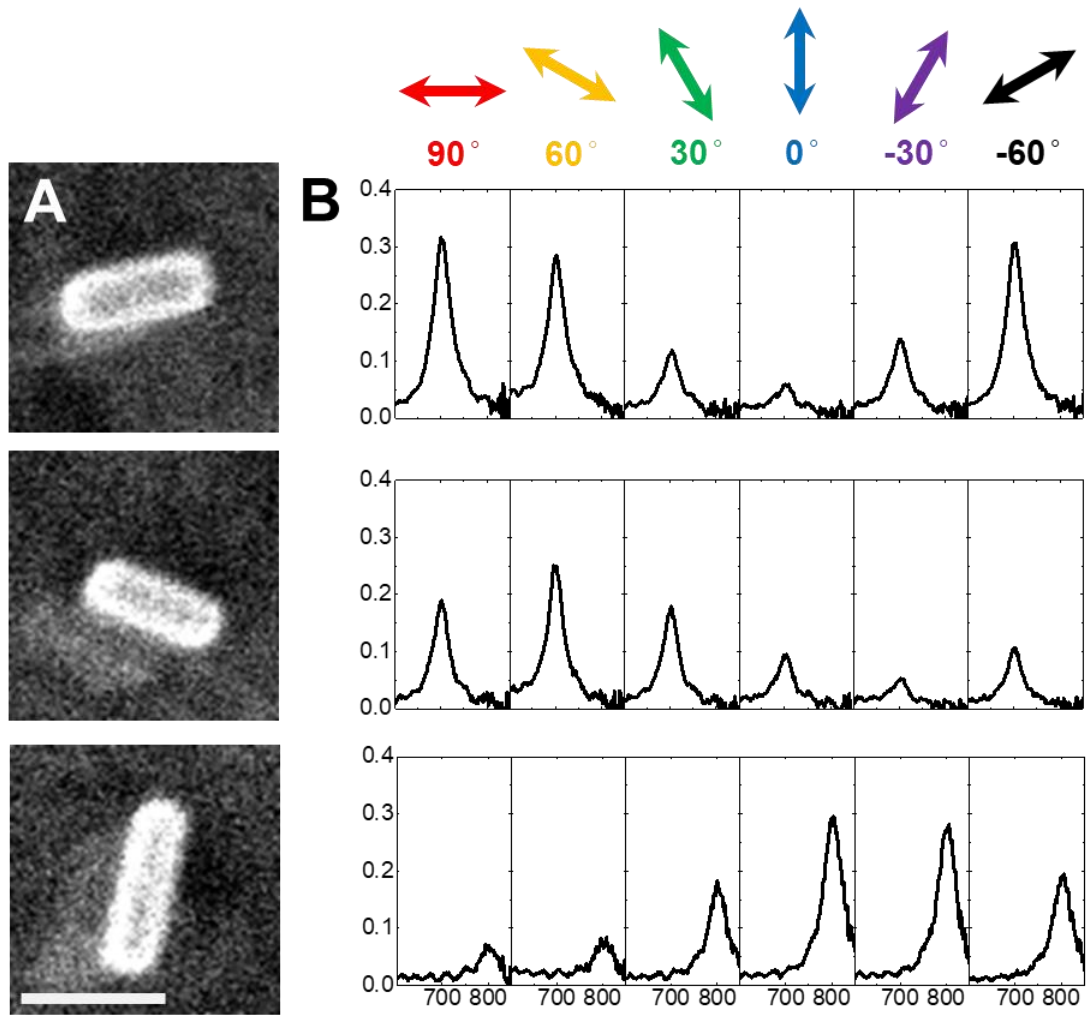

**Fig. S4.** (A) SEM images and (B) linearly polarized dark-field scattering spectra of NRs spotted in **Fig. 2**. Polarization directions were referenced and specified.

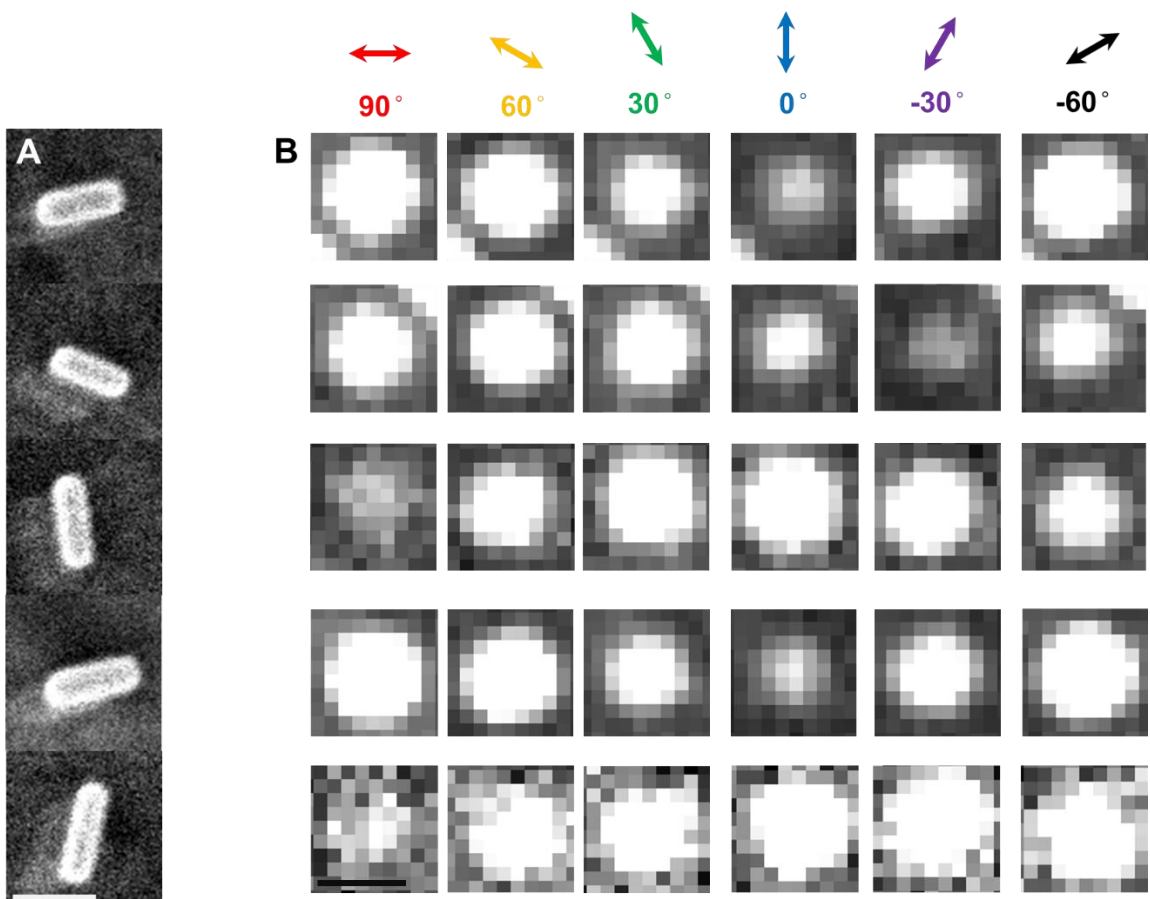

**Fig. S5.** (A) SEM images and (B) linearly polarized dark-field scattering imaging of NRs. Polarization directions were referenced and specified.

```python
import cv2
from sklearn.decomposition import PCA

def compute_orientation(image):
    """Computes the primary orientation of an image using PCA."""
    # Convert to grayscale if needed
    if len(image.shape) > 2:
        image = cv2.cvtColor(image, cv2.COLOR_BGR2GRAY)

    # Threshold the image to obtain binary mask
    _, binary = cv2.threshold(image, 0, 255, cv2.THRESH_BINARY + cv2.THRESH_OTSU)

    # Find nonzero pixel coordinates
    coords = np.column_stack(np.where(binary > 0))

    # Apply PCA=-
    pca = PCA(n_components=2)
    pca.fit(coords)

    # Principal axis (first component)
    eigenvector = pca.components_[0]
    angle = np.arctan2(eigenvector[1], eigenvector[0]) * 180 / np.pi

    if angle > 90:
      angle = angle - 180
    return angle

# Process each image individually
angles = np.array([compute_orientation(img) for img in im_stack])
```

**Fig. S6**. Algorithm snippet of applying Principal Component Analysis (PCA) to calculate in-plane orientational angle of NRs. This method converts the image to grayscale format, extracts coordinate information, identifies the principal axis and finally performs an inverse tangent operation to calculate the orientational angle relative to x-axis. We also rescale the angle referencing.

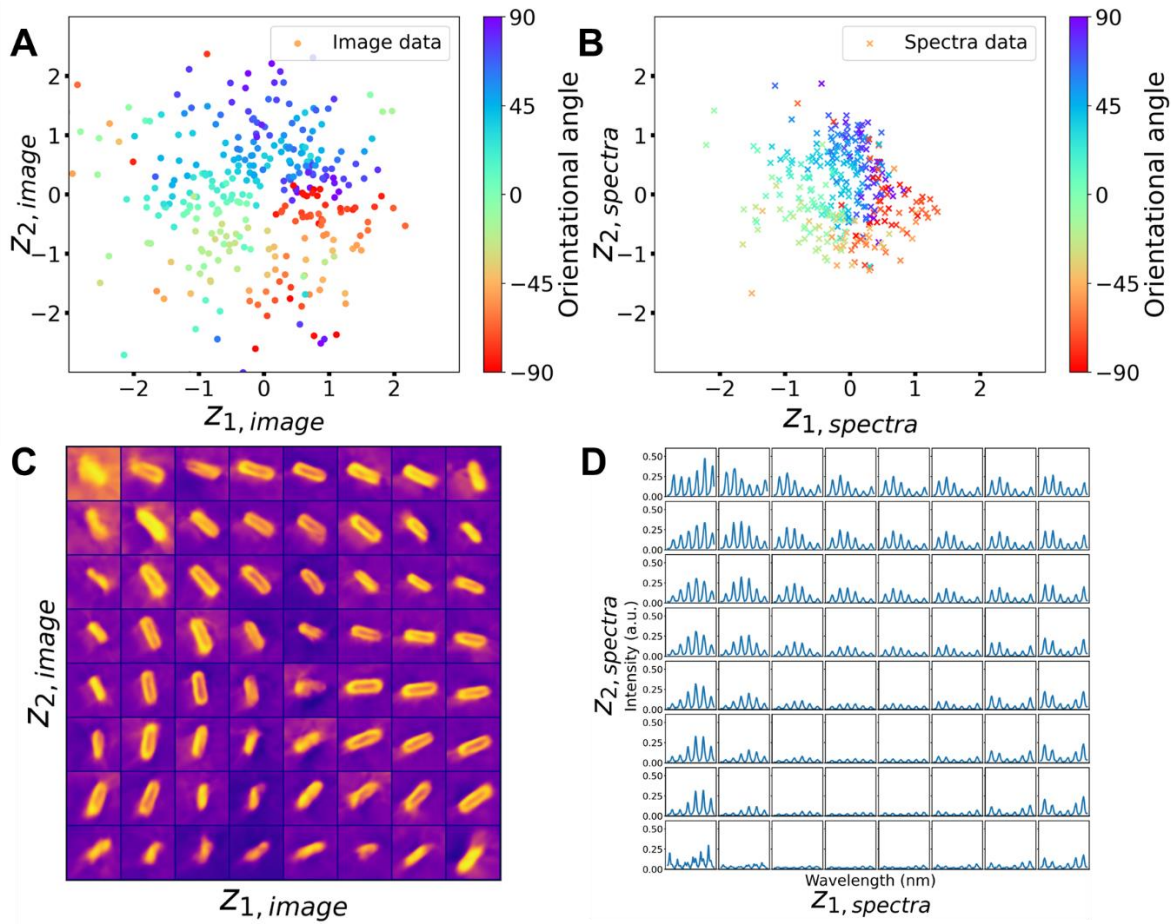

**Fig. S7.** Latent space of dual-VAE with two latent vectors. Scatterplots of encoded latent variables for image VAE (A) and spectra VAE (B). Manifold representation of decoded images (C) and decoded spectra (D).

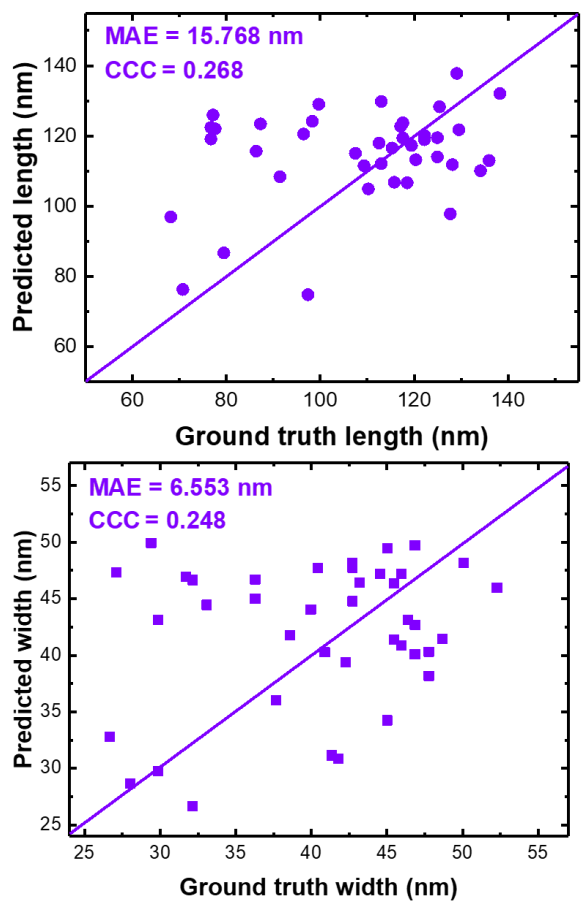

**Fig. S8.** Size prediction performance of dual-VAE with 2D latent space. Mean absolution error (MAE) and concordance correlation coefficient (CCC) both show poor prediction performance. We calculate MAE via equation (**S1**) and CCC via equation (**S2**) as shown above.

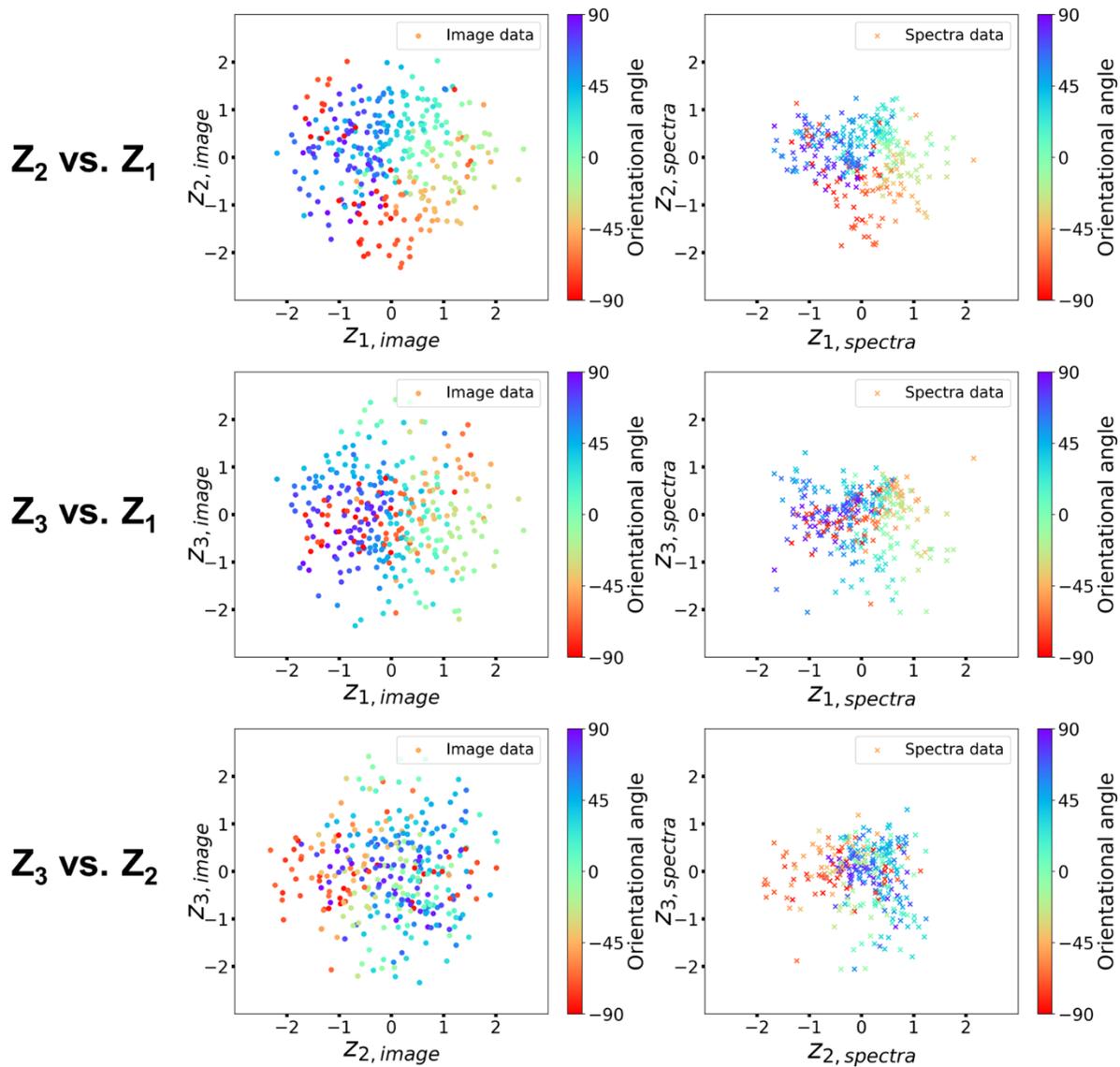

**Fig. S9.** Scatterplots of encoded latent variables of $Z_1$ vs. $Z_2$, $Z_1$ vs. $Z_3$, and $Z_2$ vs. $Z_3$ for dual-VAE with 3D latent space.

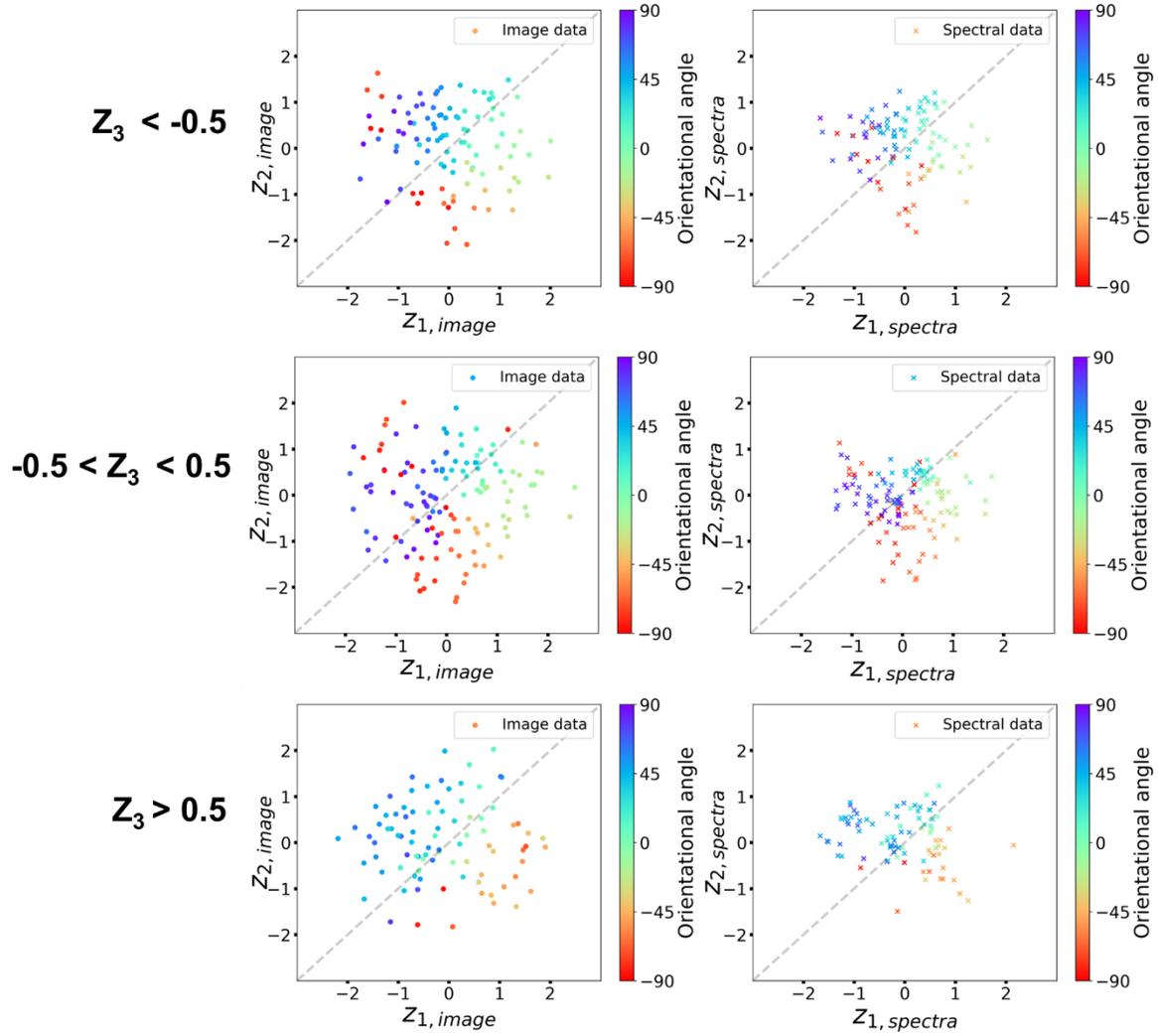

**Fig. S10.** Scatterplots of encoded latent variables of $Z_1$ vs. $Z_2$ with a fixed range of $Z_3$. A diagonal line in grey dash line separates the [-90°, 0°] and [0°, 90°] angled NRs.

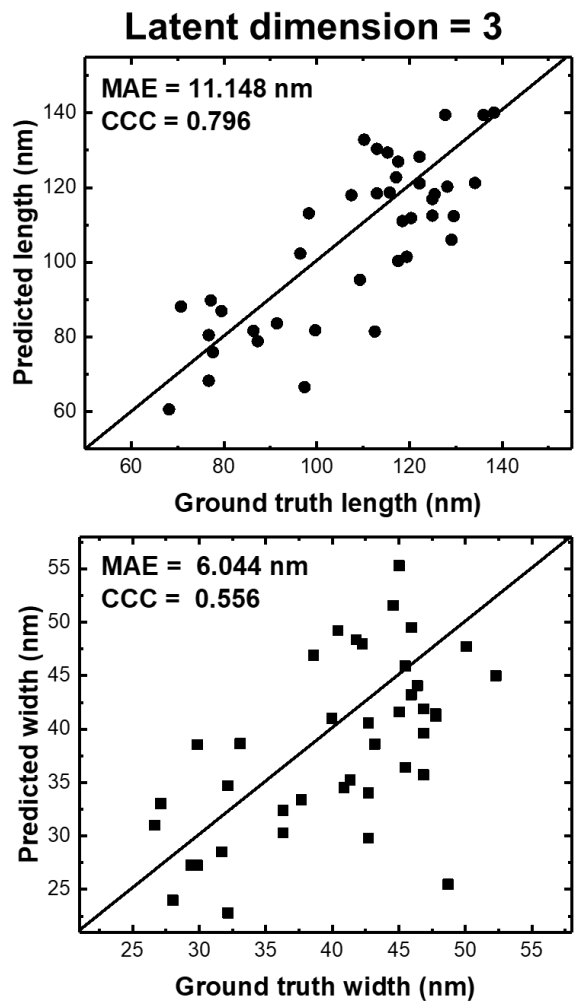

**Fig. S11.** Size prediction performance of dual-VAE with 3D latent space. MAE (for length 11.15 nm *vs.* 15.77 nm; for width, 6.04 nm *vs.* 6.55 nm) and CCC (for length, 0.80 *vs.* 0.27; for width, 0.56 *vs.* 0.25) both show improved prediction performance compared to dual-VAE with 2D latent space.

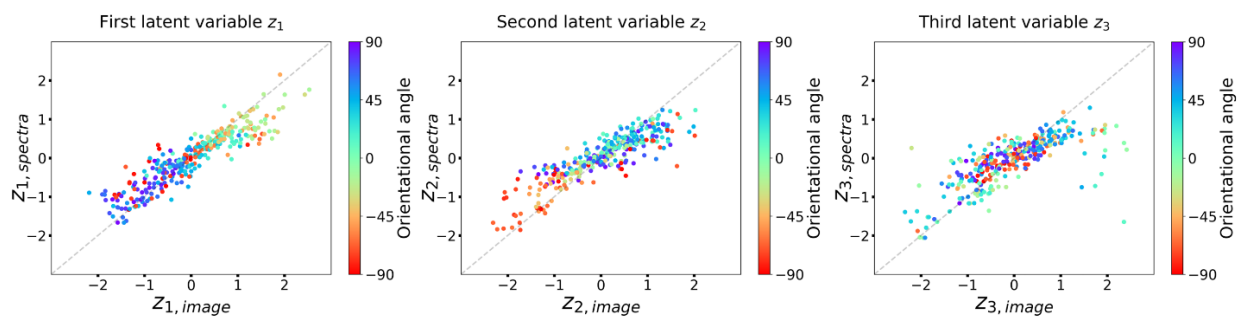

**Fig. S12.** Scatterplots of encoded latent variables across image and spectra VAE.

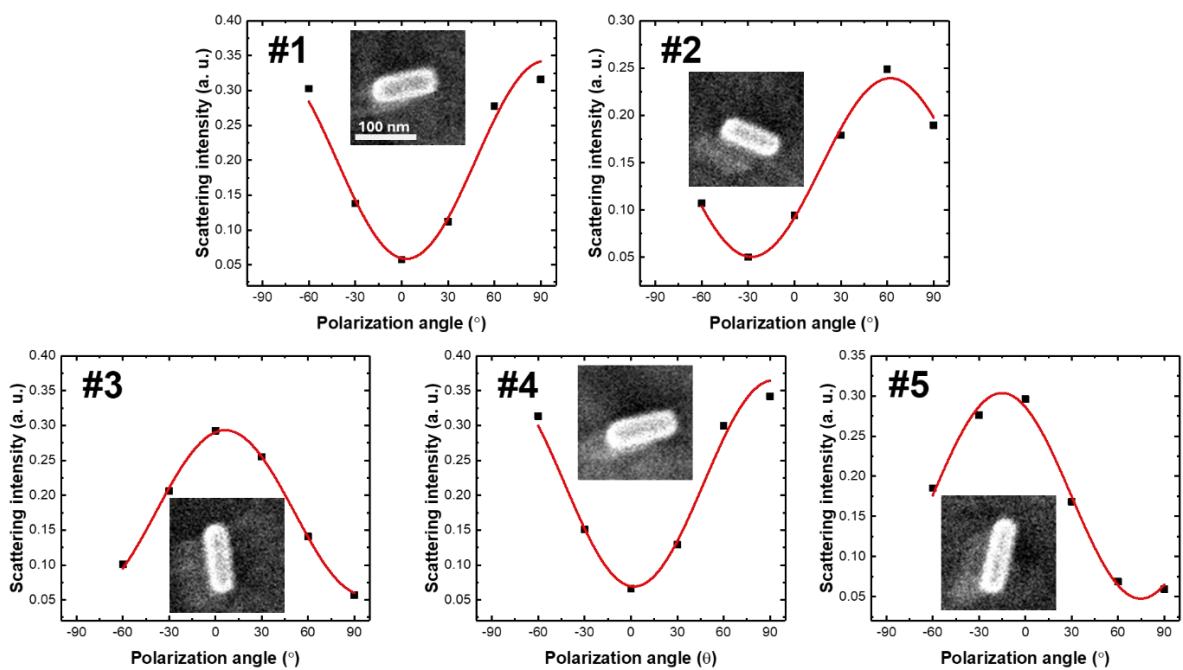

**Fig. S13.** Representative $\cos(2\theta)$ fittings of NRs spotted in **Fig. 2**. SEM images of NRs are included in the inset. The scale bar is 100 nm.

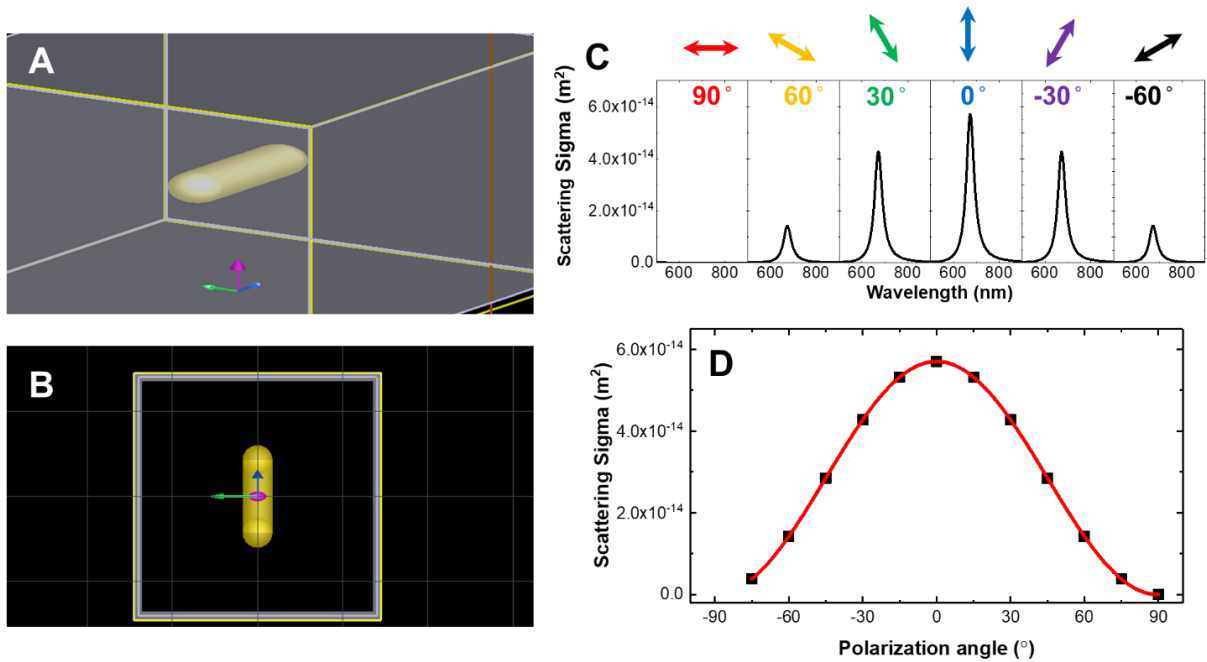

**Fig. S14.** FDTD simulation for orientation dependence of scattering of single nanorod. (A) perspective view and (B) light source view of the simulation box. Purple vector denotes the light propagation direction, blue vector denotes the electronic field vector, and green vector denotes the magnetic field vector. (C) Linearly polarized scattering cross-section patterns. We specify polarization directions for each pattern. (D) Cos(2$\theta$) fitting curves.

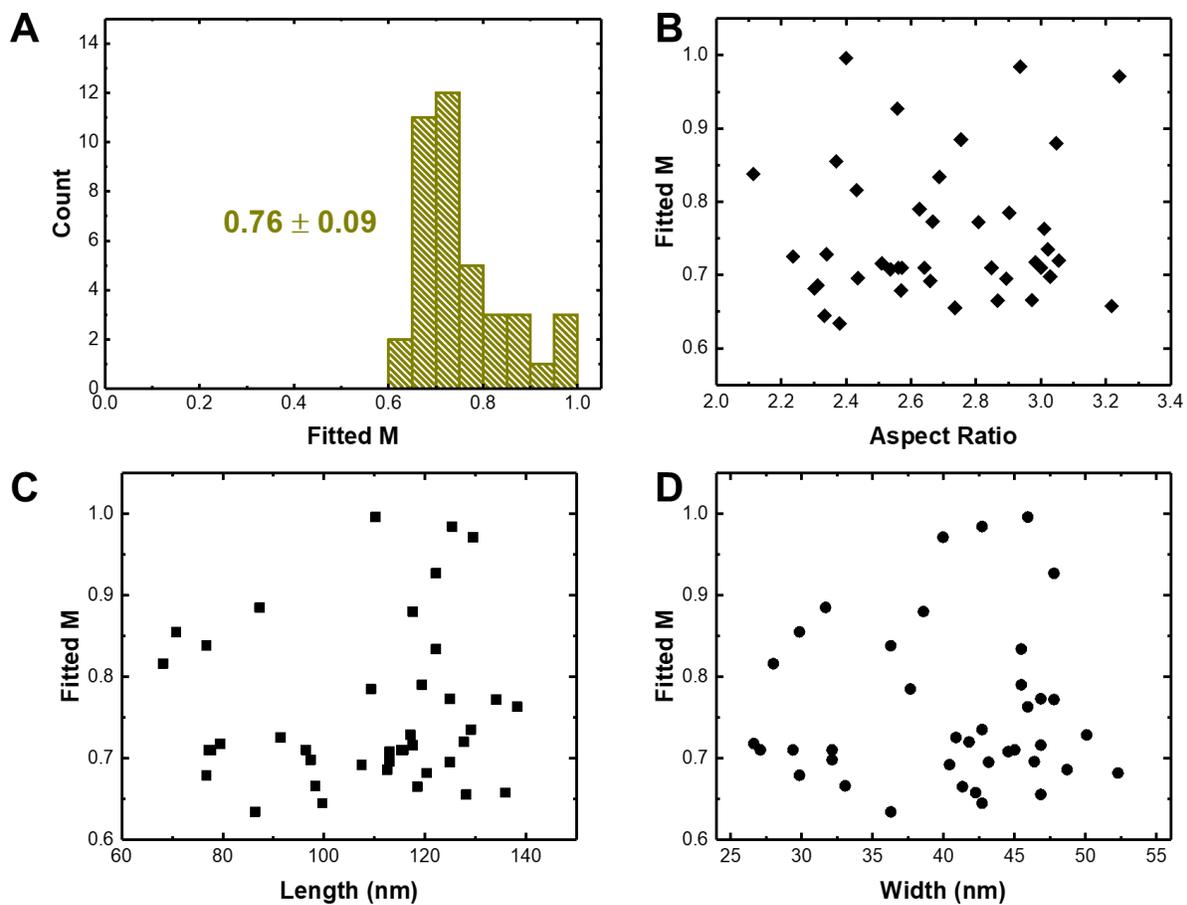

**Fig. S15.** (A) Histogram of fitted modulation depth and scatterplots of fitted modulation depth against (B) aspect ratio, (C) length, and (D) width of NRs.

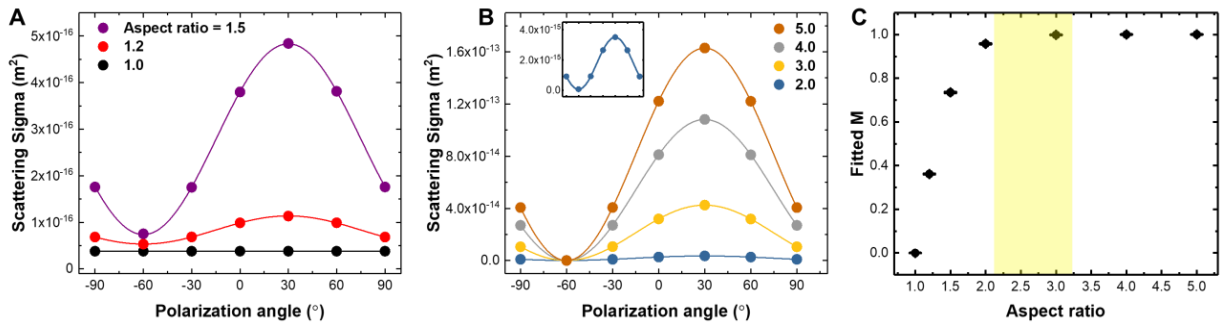

**Fig. S16.** Cos(2$\theta$) fitting curves for simulated NRs with aspect ratio of (A) 1.0, 1.2, and 1.5, and (B) 2.0, 3.0, 4.0, and 5.0. A magnified view of fitting curve of NR with aspect ratio of 2.0 is shown in the inset. (C) A scatterplot of fitted *M vs.* aspect ratio. In the simulation, we fix the width of NRs to be 40 nm and change the length to tune aspect ratio. Our testing dataset includes NRs with aspect ratio from 2.1 to 3.2, highlighted in yellow.

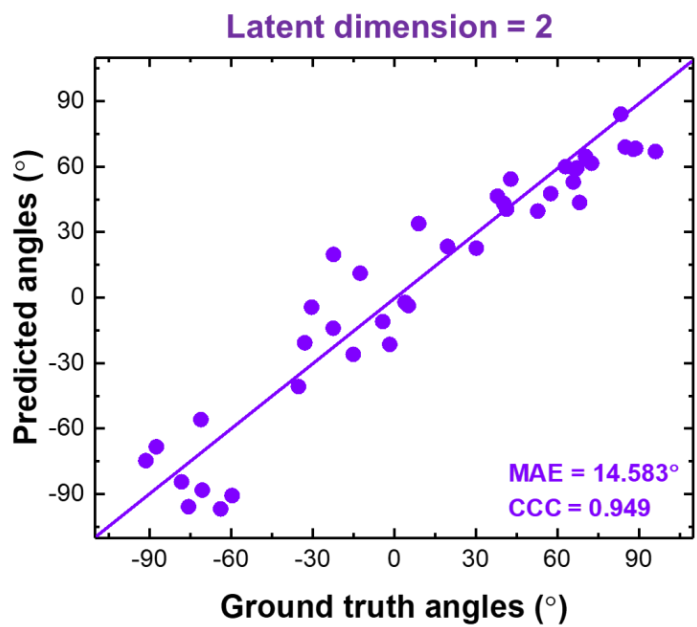

**Fig. S17.** Angle prediction performance of dual-VAE model with 2D VAE.

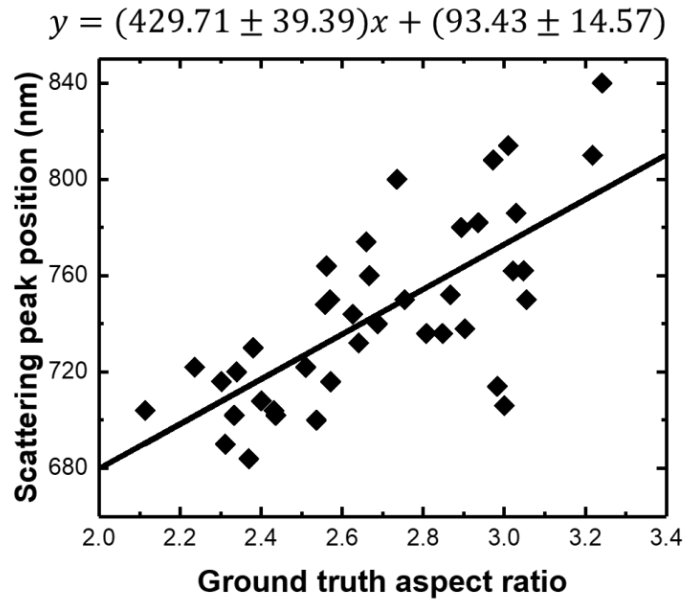

**Fig. S18.** Scatterplot and fitted line of dark-field scattering peak position against the ground truth aspect ratio of NRs in the test dataset. The equation for the best-fit line is used to determine the fitted aspect ratio in **Fig. 5**.

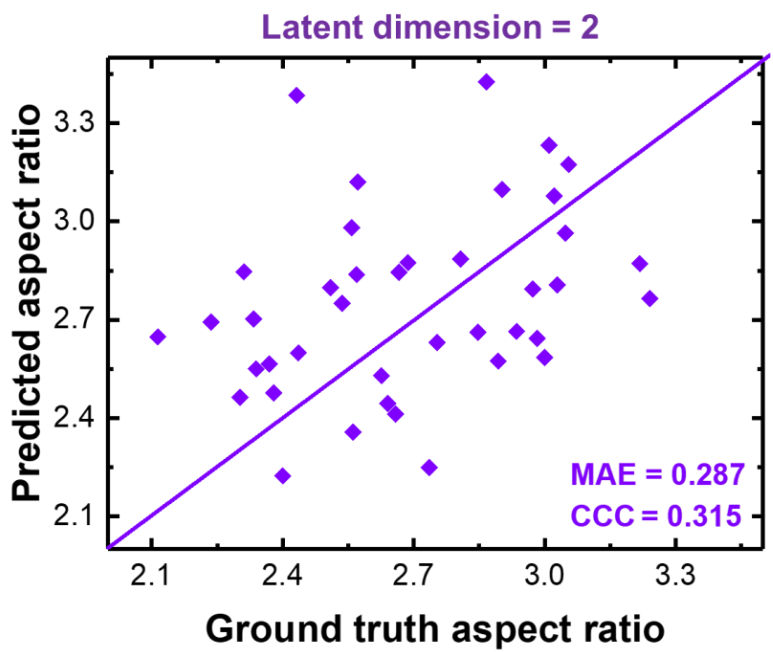

**Fig. S19.** Aspect ratio prediction performance of aspect ratio via 2D dual-VAE model. We attribute the excellent angle learning performance from both 2D and 3D latent space models to the side-by-side concatenation arrangement of polarized spectra that highlights one important spectra signature that is the relative scattering intensity changes in response to incident linearly polarized light.

However, this arrangement undervalues other spectra signatures such as changes in peak position which is an important spectral feature that relates to aspect ratio of NRs. Therefore, as the dimension of the latent space reduces, in-plane orientational angle learning is maintained at high accuracy, but size prediction is impacted greatly.